\documentclass{iaus}
\usepackage{graphicx}

\title[Properties of SS433 and ULXs] 
{Properties of SS433 and ultraluminous X-ray sources in external galaxies}

\author[Fabrika, Karpov, Abolmasov \and Sholukhova]   
{S. Fabrika,  
S. Karpov, P. Abolmasov \break \and O. Sholukhova}

\affiliation{Special Astrophysical Observatory of the Russian AS, Nizhnij
           Arkhyz 369167,  Russia
\break email: fabrika@sao.ru, karpov@sao.ru, pasha@sao.ru, olga@sao.ru\\[\affilskip]}

\pubyear{2004}
\volume{230}  
\date{?? and in revised form ??}
\jname{Proceedings Title IAU Symposium}
\editors{E.J.A. Meurs \& G. Fabbiano, eds.}
\begin{document}

\maketitle

\begin{abstract}
We suggest that the ultraluminous X-ray sources located in external galaxies
(ULXs) are supercritical accretion disks like that in SS433, observed close
to the disk axis. We estimate parameters of the SS433 funnel, where the
relativistic jets are formed. Emergent X-ray spectrum in the proposed model of 
the multicolor funnel (MCF) is calculated. The spectrum can be compared 
with those of ULXs. We predict a complex absorption-line spectrum with broad 
and shallow K$\alpha$/Kc blends of the most abundant heavy 
elements and particular temporal variability. Another critical idea 
comes from observations of
nebulae around the ULXs. We present results of 3D-spectroscopy of nebulae
of two ULXs located in Holmberg II and NGC6946. In both cases the 
nebula is found to be powered by the central black hole. The nebulae are 
compared with SS433 nebula (W50).
\end{abstract}

\firstsection 
\section{Introduction}

The main properties of the ultraluminous X-ray sources (ULXs)~-- 
huge luminosities ($10^{39-41}\,erg/s$), diversity of X-ray spectra,
strong variability, connection with star-forming regions, their surrounding
nebulae, were reviewed by \cite{Ward05}. ULXs may be supercritical accretion 
disks observed close to the disk axis in close binaries with a stellar mass
black hole or microquasars (\cite{FaMe01}; \cite{Kingetal01}; 
\cite{Koerdetal01}), or 
they may be intermediate-mass black holes with "normal" accretion disks
(\cite{CoMu99}; \cite{Milletal04}). It is also possible that ULXs 
are not homogeneous class of objects. It was suggested originally by 
\cite{Katz87} that SS433 being observed close to the jet axis will be
extremely bright X-ray source. \cite{FaMe01} discussed  observational 
properties of face-on SS433-like objects and concluded that they may appear 
as a new type of extragalactic X-ray sources. Here we continue to develop
this idea. 

\section{The funnel in SS433 disk}

The main difference between SS433 and other known X-ray binaries is
highly supercritical and persistent mass accretion rate 
($\sim 10^{-4}\,M_{\odot}/y$) onto the relativistic star (a probable 
black hole, $\sim 10 M_{\odot}$), which has led to the formation of a 
supercritical accretion disk and the relativistic jets. SS433 properties
were reviewed recently by \cite{Fab04}.

SS433 is an edge-on system, its total luminosity
(mainly in UV) is $L_{bol} \sim 10^{40}\,erg/s$. A temperature of the source
is $T = (5-7)\cdot10^4\,K$, a  
mass loss rate in the wind is $\dot M_w \sim 10^{-4}\,M_{\odot}/y$.
Both optical and X-ray jets are well collimated ($\sim 1^{\circ}$).  
SS433 X-ray luminosity (the cooling X-ray jets) 
is $\sim 10^4$ times less than 
bolometric luminosity, however kinetic luminosity of the jets 
is very high, $L_{k} \sim 10^{39}\,erg/s$.  

The jets have to be formed in a funnel in the disk and the disk wind. 
Supercritical  
accretion disks simulations (\cite{Egetal85}; \cite{Okuda02}; 
\cite{Okudaetal05}) show that a wide funnel is formed (a full opening 
angle $\theta_f \approx 40^{\circ}-50^{\circ}$) close to the black hole.
Convection is important in the inner accretion disk outside the 
funnel walls.  

If one adopts for the funnel luminosity, that it is about the same
as SS433 bolometric luminosity, $L_f \approx L_{bol}$, 
(\cite{FaMe01}; \cite{FaKa05}) and  
the funnel opening angle $\theta_f \approx 40^{\circ}-50^{\circ}$, 
one obtains "observed" face-on luminosity of SS433 
$L_x \sim 10^{41}\,erg/s$ and expected frequency of such objects 
$\sim 0.1$ per a galaxy like Milky Way. On the other hand a critical luminosity    
for a $10\,M_{\odot}$ black hole is $L_{edd} \sim 10^{39}\,erg/s$.  
At highly supercritical accretion rate $\dot M / \dot M_{cr} \sim 10^3$ 
it increases (the inner disk geometry) by a factor of
$(1 + \ln(\dot M / \dot M_{cr})) \sim 10$. A doppler boosting factor is not 
large for SS433 ($\beta=V_j/c=0.26$), it is 
$1 / (1-\beta)^{2+\alpha} \sim 2.5$. The third factor is the 
geometrical funneling ($\theta_f \approx 40^{\circ}-50^{\circ}$), it is 
$\Omega_f/2\pi \sim 10$. Thus, one may expect a face-on luminosity 
of such supercritical disk $L_{edd} \sim 2 \cdot 10^{41}\,erg/s$.

\section{The multicolor funnel model}

A size of SS433 wind photosphere is $r_{ph, wind} \sim  10^{12}\,cm$. 
An estimate of the jet photosphere (\cite{FaKa05}), which is a bottom of
the wide funnel, gives $r_{ph, jet} \sim 4 \cdot 10^9\,cm$.
We find that the jet is not transparent down to the black hole horizon. 
However, the gas supply to the funnel is very variable due to
convection (\cite{Okudaetal05}).

\cite{FaKa05} developed a simple model of the funnel 
(multicolor funnel, MCF) to estimate the emerging X-ray spectrum. 
They considered both gas pressure dominated, $T(r) \propto r^{-1}$, and
radiation pressure dominated, $T(r) \propto r^{-1/2}$ cases.
They based on the observed temperature of the wind photosphere
in SS433 and found temperatures of the inner funnel walls
(at a level of $r_{ph, jet}$), $T_{ph, wind}$ between  
$\sim 1.7 \cdot 10^7\,K$ and $\sim 1 \cdot 10^6\,K$. Below 
$r_{ph, jet}$ the walls can not be observed. 

\begin{figure}
{\centering 
 \includegraphics[height=5cm,angle=-90]{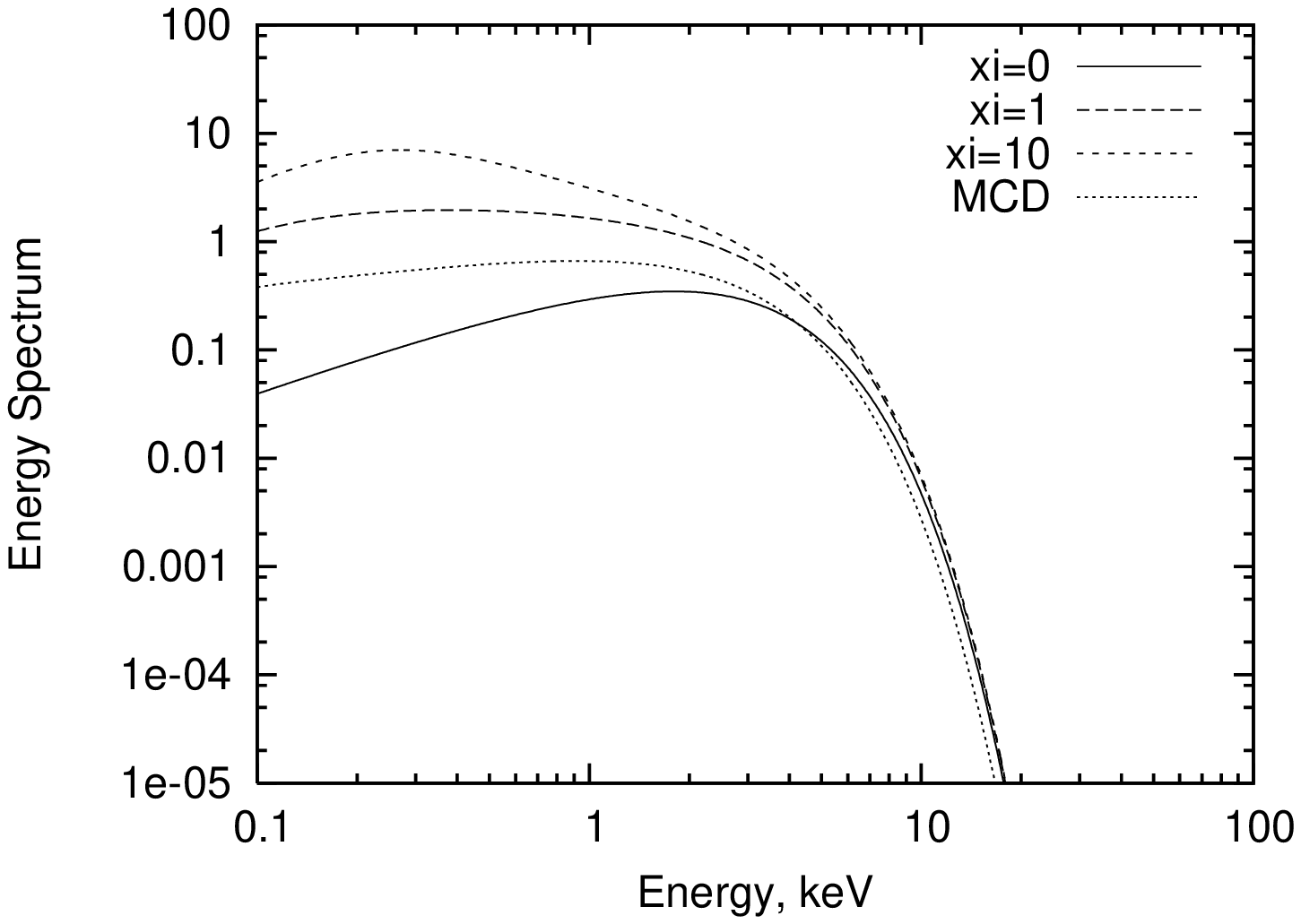}
 \includegraphics[height=5cm,angle=-90]{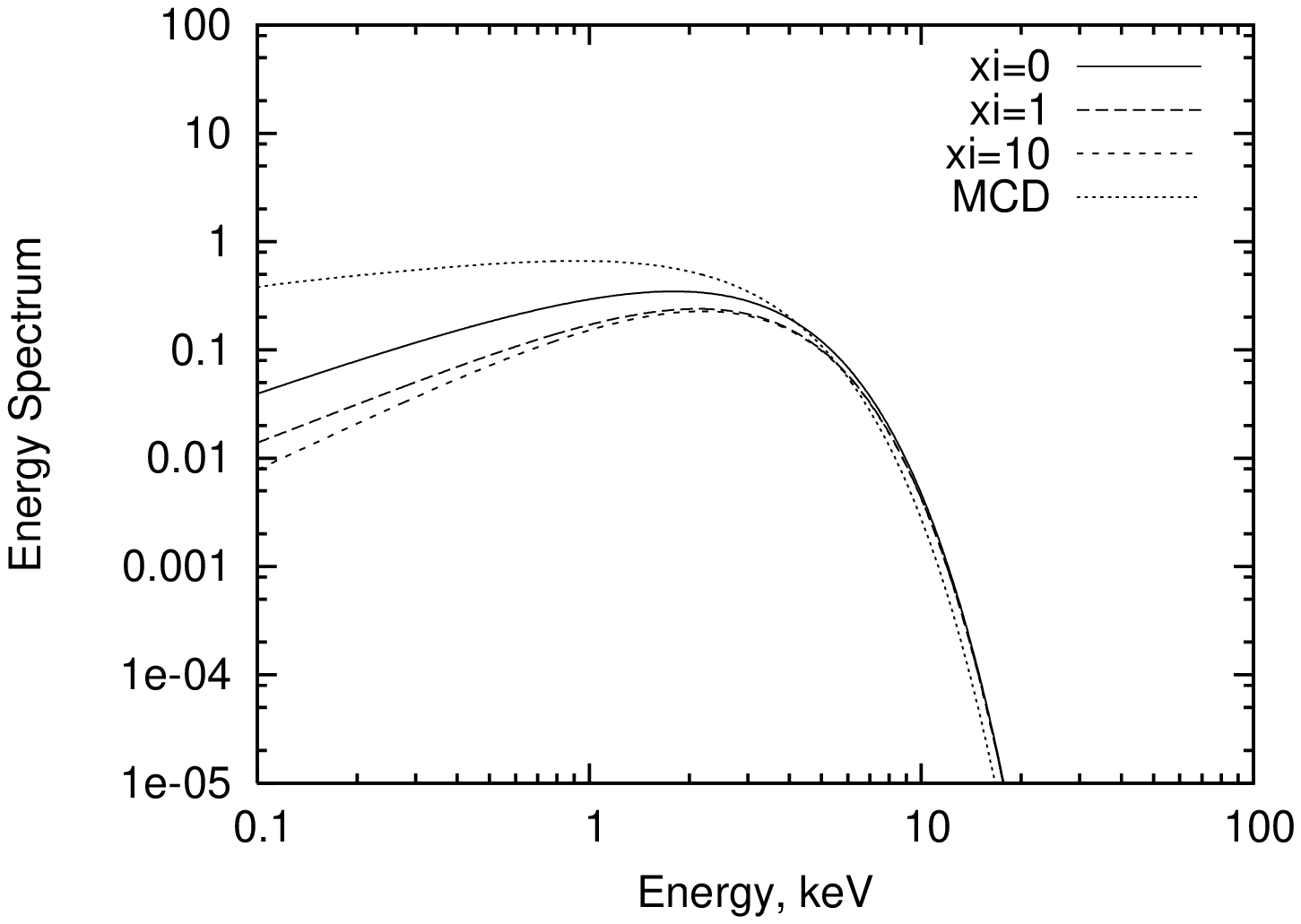}
\hspace*{0.1cm}
 \includegraphics[height=4.8cm,angle=-90]{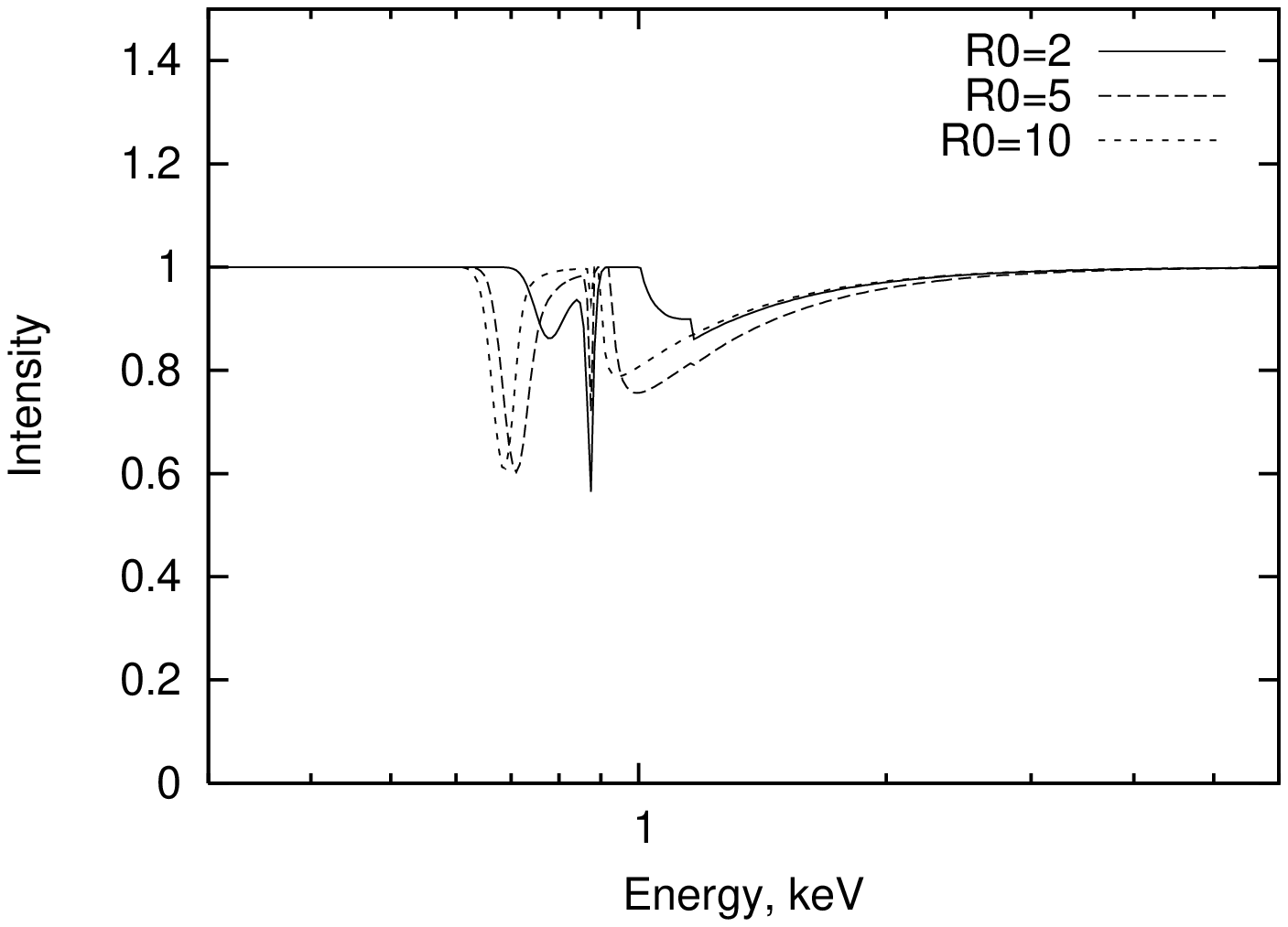}
\hspace*{0.1cm}
 \includegraphics[height=4.8cm,angle=-90]{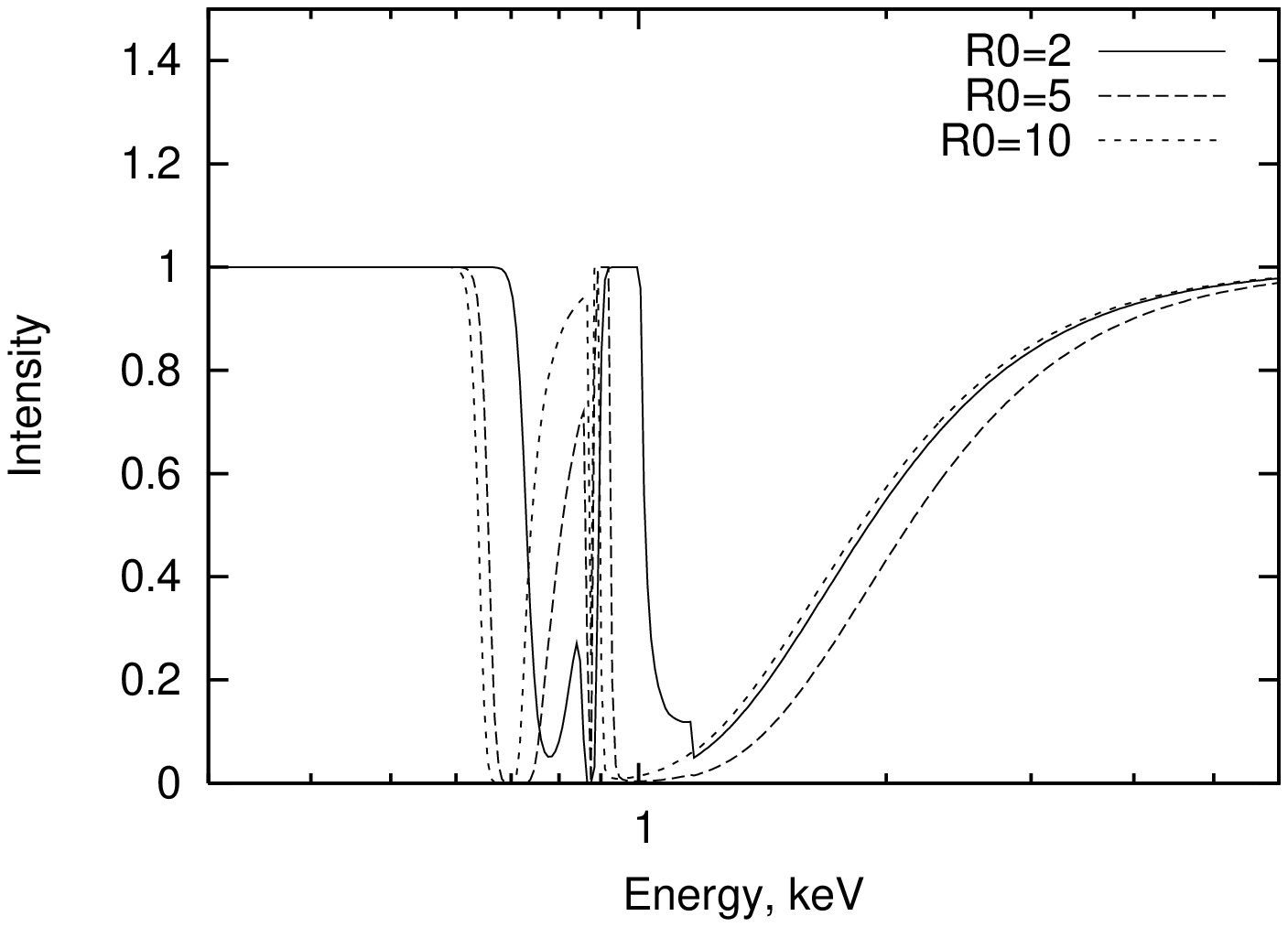}
\par}
  \caption{Top: energy spectra of the multicolor funnel (MCF) 
compared with the MCD spectrum with the same inner temperature $1\,keV$. 
Bottom: absorption line profiles of OVIII L$\alpha$ and Lc transitions 
at different
efficiencies of jet acceleration for optically thin (left) and for optically 
thick (right) regimes. The acceleration starts at $R_0=1$
(in units of the jet bottom photosphere, $r_{ph, jet}$), and ends at 
$R_0=2, 5, 10$. The line intensities are in arbitrary scale.
}
\end{figure}

In Fig\.1 we present MCF energy spectra with different ratios of radiation 
to gas pressure ($\xi={aT_0^3}/{3k_b n_0} $) at a deepest visible
parts of the funnel's walls ($r_{ph, jet}$). 
The temperature $T_0 = 1\,keV$ was adopted.
They are spectra with no radiation losses in the funnel 
($\theta_f \approx 0$), and with radiation losses included 
($\theta_f \approx 40^{\circ}$). The multicolor disk (MCD) 
spectrum with the same inner temperature $T_{inn} = 1\,keV$ 
is shown for comparison.  
A low-temperature break in the MCF spectra appears due to finite length
of the funnel.

The jets velocity ($0.26\,c$) in SS433 and its surprising stability suggest
line-locking mechanism (\cite{ShaMiRe86}) to operate in the funnel.
A photosphere of the funnel's inner walls may radiate absorption spectrum
with Lc and Kc edges of H- and He-like ions. The MCF model predicts
very complex absorption-line spectrum. Fig.\,1 presents expected
absorption line profiles of OVIII L$\alpha$ and Lc transitions.

The absorption bands should belong to H- and He-like
ions of the most abundant heavy elements and
should extend from the Kc to the K$\alpha$ energies of the corresponding
ions.
Variations of the gas parameters along the funnel~-- its velocity, density,
temperature and volume filling factor (the collimation)~-- could make 
the absorption-line profiles appreciably more complex, necessitating 
the use of X-ray spectra
with high signal/noise ratios in searches for these lines.

The supposition that ULXs are face-on SS433 objects leads to the following 
predictions:
1) We expect typical MCD-like spectra, however the MCF spectra may be 
more diverse depending on the funnel parameters. 
2) Absorption spectrum with broad and shallow L$\alpha$/Lc, K$\alpha$/Kc 
blends of the most abundant heavy elements. 
3) Temporal variability on time scales $r_{ph, wind}/c \sim 30\,sec$ and 
$r_{ph, jet}/c \sim 0.1\, sec$.
4) A typical accretion disk power density spectrum at scales $>> 0.1\, sec$.  
5) Very bright UV source ($L_{UV} \sim 10^{40}\,erg/s$, $T \ge 10^5\, K$) 
is less collimated than X-ray source. 
The predicted complexity of the dependence of the absorption-line
profiles on the structure of the funnel and mechanisms of acceleration and
collimation of the gas in the funnel presents excellent opportunities for
direct probing of these structures in supercritical accretion disks and
for studies of mechanisms of jet formation.

\section{W50 and ULXs nebulae}

The radio nebula W50 was produced (or distorted) by SS433 jets. We show
W50 in Fig.\,2 together with nebulae surrounging two ULXs in Holmberg\,II
and NGC6946 galaxies, which were studied recently by Integral-field 
spectroscopy method (\cite{Leh_ea05}; \cite{FaAbSh05}). W50 contains
bipolar nebula with optical filaments located at $\pm 0.5^{\circ}$ 
or $\pm 50\,pc$ from SS433 at places of jets termination. A total 
energy of the nebula is $E_{k} \sim 2\cdot10^{51}\, erg$ (\cite{Zealetal80}),
which corresponds to the jet kinetic luminosity $L_k \sim 3\cdot10^{39}\,erg/s$   
for 20000 years. The observed velocity dispersion in the filaments is
$\sim 50 \,km/s$, however [NII]/H$\alpha$ line ratio coresponds to
dispersion $\sim 300 \,km/s$. SS433 is an edge-on system, 
$i = 79^{\circ}$. If one takes into account this factor,
the velocity dispersion may reach $250- 300 \,km/s$.   

\begin{figure}
\includegraphics[height=4.1cm]{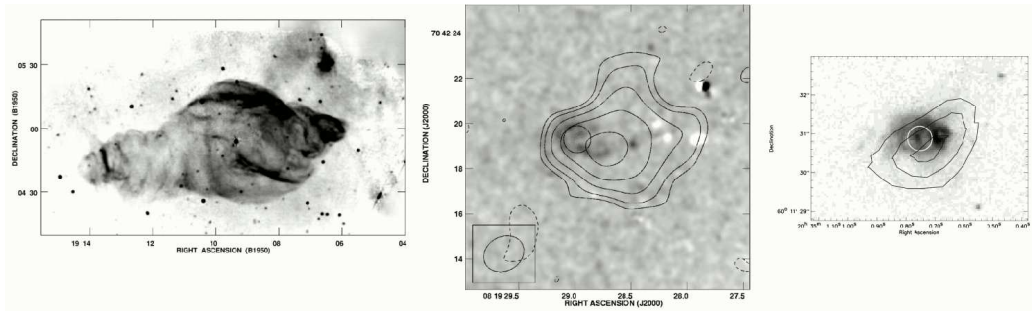}
  \caption{Three nebulae in the same scale in parsecs. VLA image of W50
(\cite{Dubetal98}) with SS433 in center, left; Holmberg\,II ULX-1
in HST HeII image with VLA isophotes, middle (\cite{MiMuNe05}) and 
NGC6946 ULX-1 in HST H$\alpha$ + [SII] image with VLA isophotes, right 
(\cite{vanDyketal94}). Circles show X-ray Chandra positions.
}
\end{figure}

ULXs are located frequently in bubble-like nebulae. New data (\cite{Pakull05})
show that the nebulae are expanding with a velocity $\sim 80 \,km/s$
(up to $\sim 250 \,km/s$). In two well-studied ULXs nebulae radial velocity
gradients in a high ionisation emission line He\,II$\lambda4686$ were found.
They are  $\pm 50\,km/s$ on spatial scale $\sim \pm 30
\,pc$ in Holmberg\,II ULX-1 (\cite{Leh_ea05}), and $\sim \pm 50\,km/s$ 
on a scale $\pm 20 \,pc$ in NGC6946 ULX-1 (\cite{FaAbSh05}). 
In all cases line ratios indicate 
shock ionisation. This means that the nebulae are powered by the central 
source, probably via jets activity.

The nebulae in Holmberg\,II and NGC6946 have circle-like features in 
the line-images. In both cases the radio sources are shifted to a brighter
circle-like feature and in the both cases this part of the nebulae is  
approaching, the opposite part is receding (\cite{Leh_ea05}; 
\cite{FaAbSh05}). At some imagination one may conclude that the nebulae
around these two ULXs are face-on versions ($i=10^{\circ} - 30^{\circ}$) 
of the SS433 nebula.

\begin{acknowledgments}
The work is supported by RFBR under grants number 03-02-16341 and 
04-02-16349.
\end{acknowledgments}

\end{document}